\documentstyle[12pt,lathuile,epsfig]{article}
\begin{document}
\title{ 
NEW RESULTS FROM K2K
}
\author{
T. Ishii (for the K2K Collaboration) \\
{\em Institute of Particle and Nuclear Studies, KEK, Tsukuba, Ibaraki
 305-0801, Japan}
}
\maketitle
\baselineskip=14.5pt
\begin{abstract}
During the first half of the K2K run, data corresponding to 
$4.8 \times 10^{19}$ POT were accumulated.  
In this data set, 56 accelerator-produced neutrino 
events were observed in the far detector, 
while the expected number from the near site measurements is 
$80.6^{+7.3}_{-8.0}$.  
This means that the probability of the null-oscillation scenario 
is limited to less than 3\%.  
A method for  long-baseline neutrino experiments has been established.  
\end{abstract}
\footnote{Talk presented at the 16th Les Rencontres 
de Physique de la Vallee d'Aoste, La Thuile, Aosta Valley, Italy, 
March 3-9, 2002}
\baselineskip=17pt
\newpage
\section{Introduction}
K2K\cite{K2K} is the first accelerator-based 
long-baseline neutrino experiment, 
of which the baseline is 250 km between the KEK 12-GeV proton synchrotron 
(KEK-PS) and the Super-Kamiokande (SK) detector.  
The atmospheric neutrino anomaly 
found by the Kamiokande experiment\cite{Kamiokande} triggered 
various studies of neutrino oscillation in the region 
$\Delta m^{2} \approx 10^{-2}$ -- $10^{-3} \rm eV^{2}$, 
where $\Delta m^{2}$ is the difference of the squared masses. 
Among them, the SK experiment\cite{SK} confirmed the deficit 
of the upward-going atmospheric $\nu_{\mu}$s and announced 
evidence for oscillation of the atmospheric neutrino.  
For the neutrino energy ($E_{\nu}$(GeV)) and flight length ($L$(km)), 
the oscillation probability can be written in terms of the 
mixing angle ($\theta$) and $\Delta m^{2}$ in a two-flavor 
approximation as follows:  
\begin{equation}
P(\nu_{\mu} \rightarrow \nu_{x}) = 
\sin^{2}2\theta \cdot \sin^{2}\frac{1.27\Delta m^{2} \cdot L}{E_{\nu}} . 
\label{eqosci} 
\end{equation}
K2K aims to firmly establish neutrino oscillation in the 
$\nu_{\mu}$-disappearance mode and in the $\nu_{e}$-appearance mode, 
with a well-defined flight length and a well-controlled and 
understood nearly pure $\nu_{\mu}$ beam.  
The average neutrino energy is 1.3 GeV and the beam composition is 
98.2\%$\nu_{\mu}$, 1.3\%$\nu_{e}$ and 0.5\%$\bar{\nu}_{\mu}$.  
\section{Detectors}
At the near site, 12-GeV protons are extracted from the KEK-PS in 
a 1.1-$\mu$sec spill every 2.2 seconds.  The extracted protons are 
bent to the Kamioka direction and injected into an Al target 
of a 30 mm diameter and 66 cm length.  
A pair of horn magnets focuses produced positive pions, 
which subsequently decay into muons and $\nu_{\mu}$s.  \\
A pion monitor is occasionally put into 
the beam line just after the horn magnets.  
The pion monitor is a gas-Cherenkov detector, which measures 
the momentum and angular distributions of pions.  
From this measurement, it gives the far/near ratio of the neutrino flux 
as a function of the neutrino energy.  A measurement is made 
when the beam parameters are changed.\cite{pimon}  \\
A muon monitor is located after a 200 m decay pipe and a beam dump.  
The muon monitor consists of a segmented ionization chamber and 
a silicon pad detector array.  It provides spill-by-spill monitoring 
of the intensity and direction of the beam.  \\
A set of near neutrino detectors is located 300 m from the target.  
It consists of a 1kton-water-Cherenkov detector (1kt), 
a scintillating fiber tracker with a water target (SCIFI),\cite{SCIFI} 
lead-glass counters and 
a muon range detector (MRD).\cite{MRD}  
The 1kt is used to measure the neutrino flux at the near site 
to normalize the event rate at the far site while taking advantage of being 
the same type detector as the SK.  
Since the MRD is massive and has large-area coverage, 
it is used to monitor the neutrino beam.  
The SCIFI together with the MRD is used to study 
neutrino interactions.  \\
K2K uses the SK as a far neutrino detector.  
It is a 50 kton water-Cherenkov detector comprising 11 thousand inner 
PMTs and 18 hundred outer PMTs.  
\section{New results}
\subsection{Delivered beam}
\begin{figure}[bhtp]
\centerline{\epsfig{file=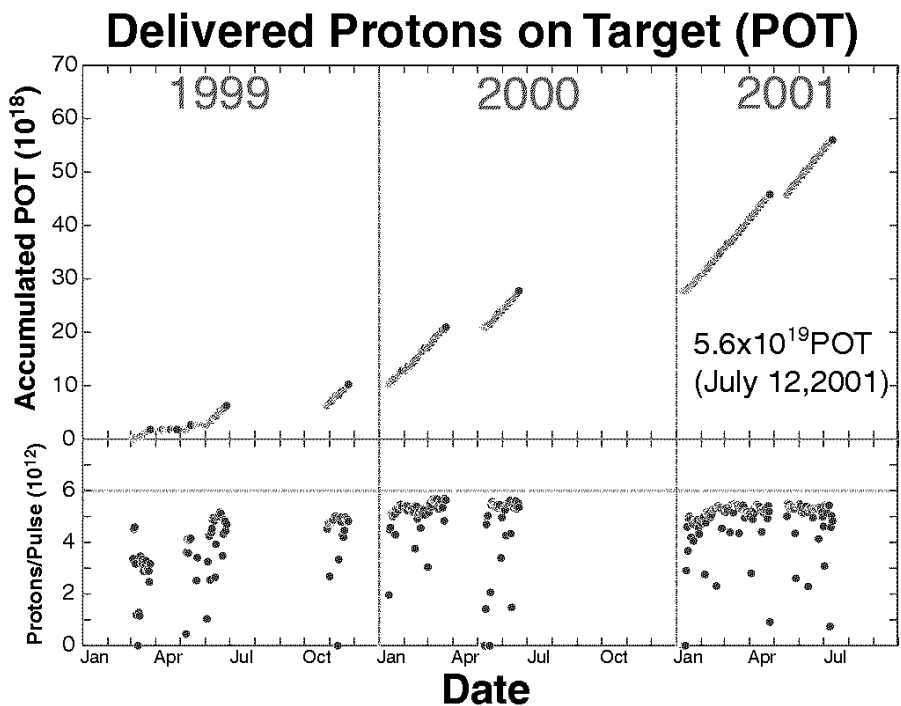,width=12cm}}
 \caption{\it
      (top) Integrated protons on the target; (bottom) Instantaneous 
       protons on the target.
    \label{intpot} }
\end{figure}

Fig.~\ref{intpot} shows the delivered protons on the target (POT) 
as a function of time.  
The top plot shows the integrated POT and the bottom plot shows 
the instantaneous POT.  The instantaneous POT has nearly reached 
its design value of $6 \times 10^{12}$ protons/pulse.  
Data corresponding to 
$4.8 \times 10^{19}$ POT are used for analyses, which were accumulated 
during the period between June 1999 and July 2001.  
The goal is to accumulate $10^{20}$ POT for analyses.  
%
%
\subsection{Beam stability}
\begin{figure}[bhtp]
\centerline{\epsfig{file=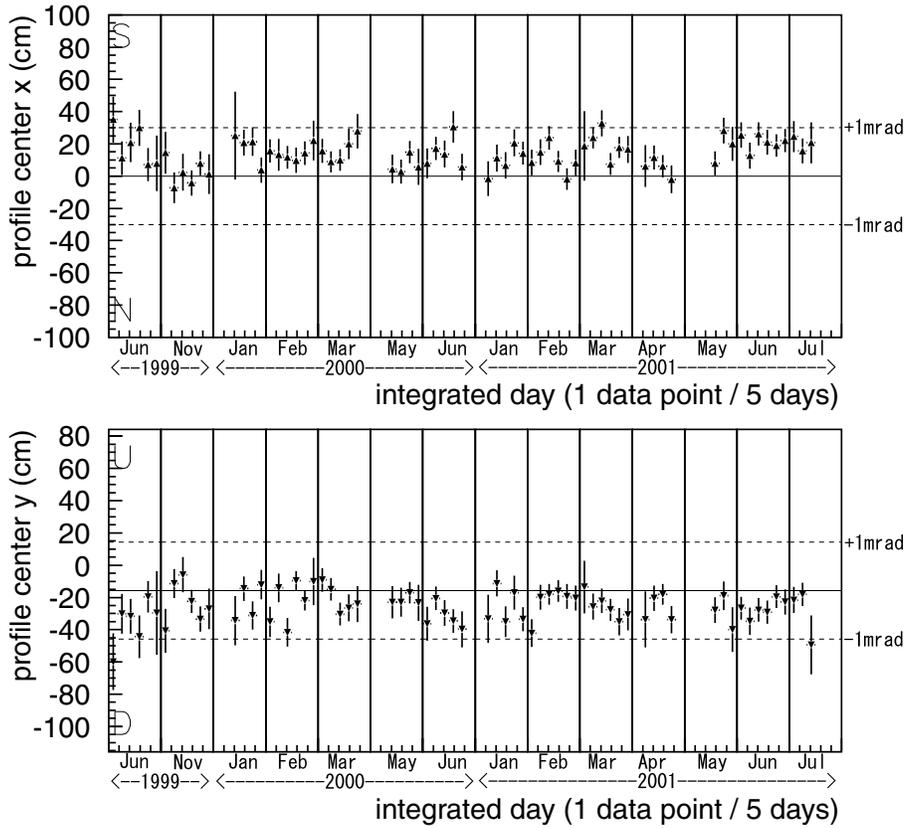,width=12cm}}
 \caption{\it
      Stability of the neutrino profile center.
      The dashed lines show $\pm 1m$rad directions to SK. 
      (top) Horizontal direction; 
      (bottom) Vertical direction. 
    \label{profstab} }
\end{figure}

\begin{figure}[bhtp]
\centerline{\epsfig{file=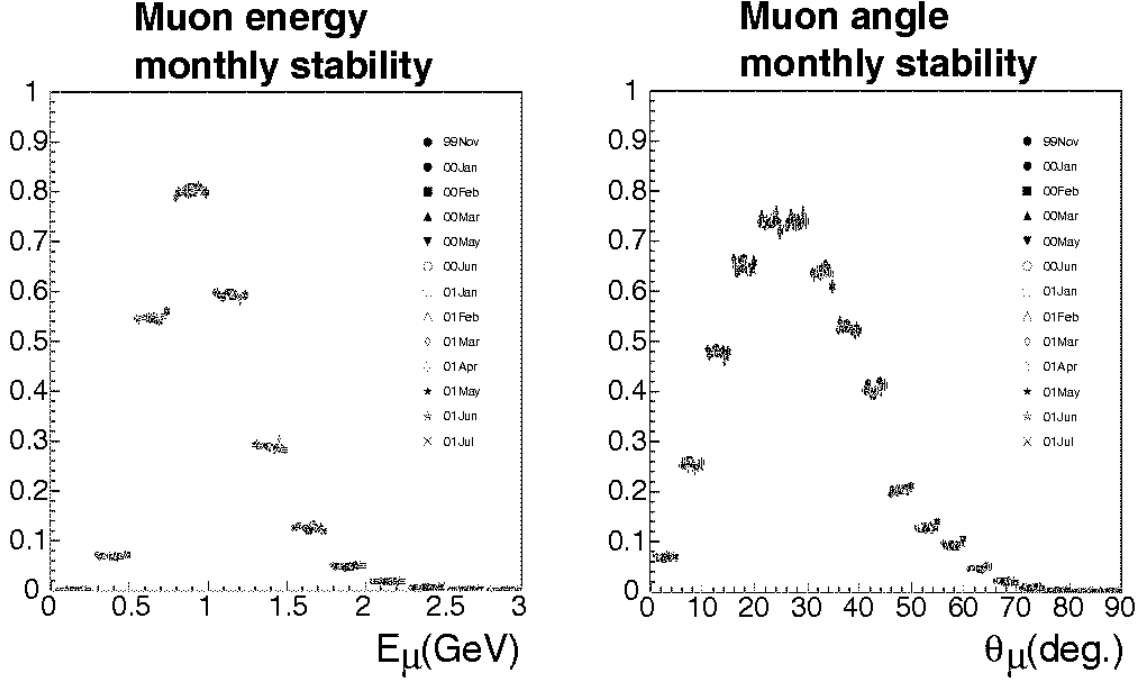,width=15cm}}
 \caption{\it
      (left) Muon energy distribution plotted each month; 
      (right) Muon angular distribution plotted each month. 
    \label{muethstab} }
\end{figure}

The neutrino beam direction has been monitored by the MRD 
by watching the neutrino profile center using $\nu_{\mu}$ 
interactions with iron.  As shown in fig.~\ref{profstab}, 
the neutrino beam has been directed to the SK detector 
within $\pm 1m$rad.  It has also been confirmed pulse-by-pulse by the muon 
monitor by watching the muon profile center.  This is sufficient 
direction stability for the experiment.  
The MRD has also monitored the neutrino event rate, and shows 
good stability.  
Fig.~\ref{muethstab} shows the energy and angular distributions 
of muons produced by the charged-current $\nu_{\mu}$ 
interaction, which were measured by the MRD 
and plotted monthly.  Both energy and angular distributions show 
good stability, which implies that the neutrino spectrum has been stable.  
%
%
%
%
\subsection{SK events}
\begin{figure}[bhtp]
\centerline{\epsfig{file=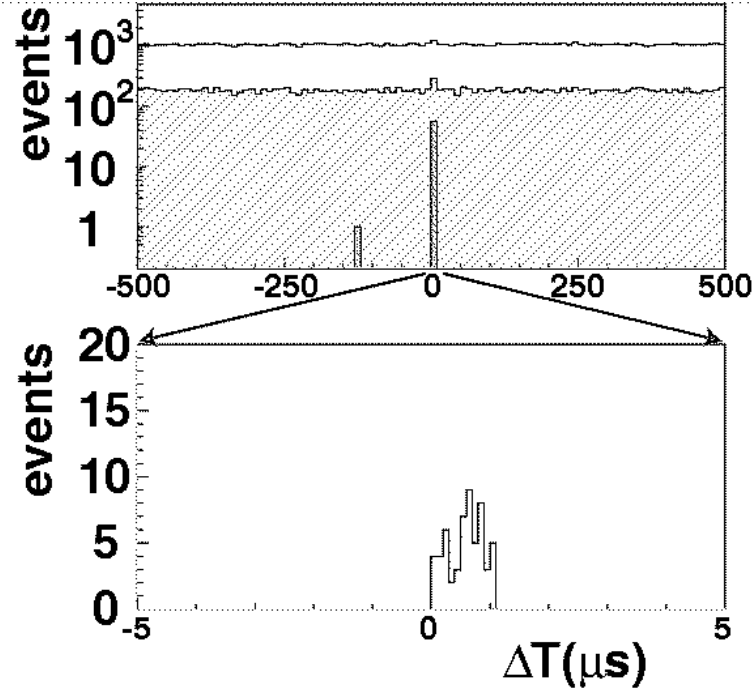,width=10cm}}
 \caption{\it
    (top) Time-difference distribution between the beam-spill start 
      and the SK event occurrence after each selection cut: 
      (unfilled histogram) no detector pre-activity, 
      (hatched) more than 200 photoelectrons in 300nsec and 
      (shaded) all cuts except timing; 
    (bottom) Time-difference distribution after all cuts in 
      a $\pm 5\mu$sec range.  
    \label{tdiff} }
\end{figure}

\begin{figure}[bhtp]
\epsfig{file=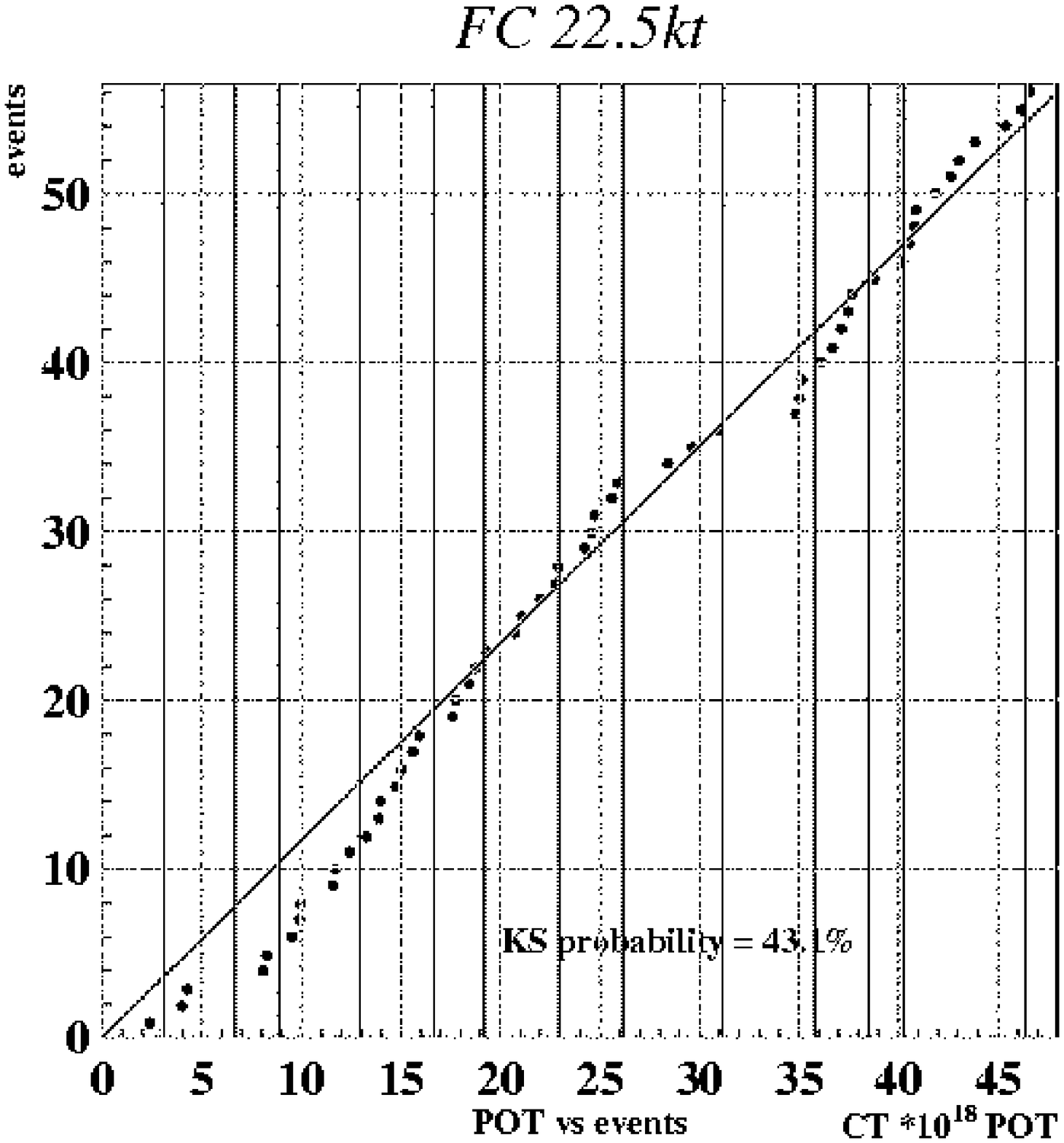,width=7cm}
\hspace{0.5cm}
\epsfig{file=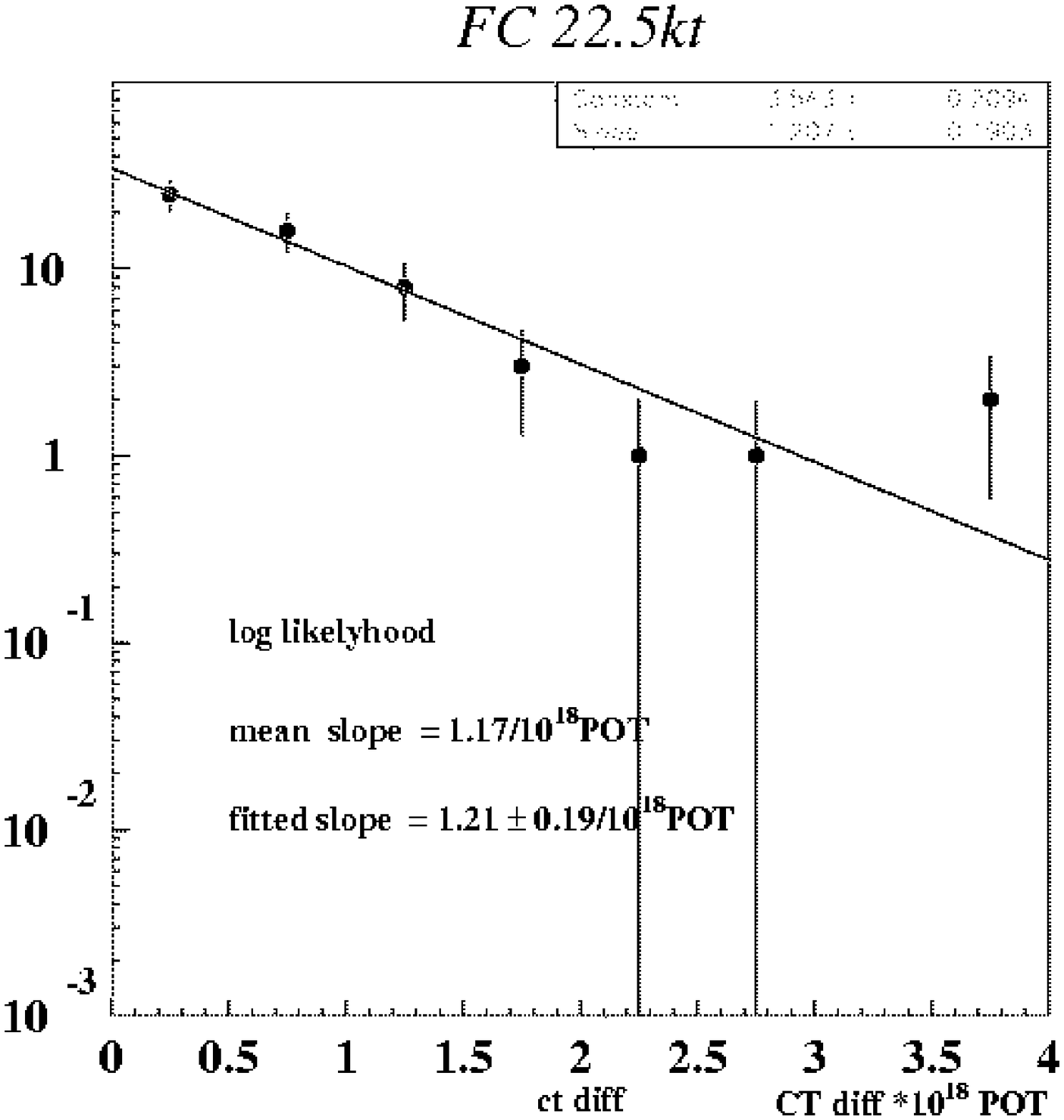,width=7cm}
 \caption{\it
      (left) SK event number vs. POT; 
      (right) Interval distribution between two consecutive events. 
        The line shows the result of the exponential fit.
    \label{SKevPOT} }
\end{figure}

Accelerator-produced neutrino events at the SK detector are 
selected essentially using the timing information of the 
global positioning system (GPS).  
Other selection cuts are the same as the atmospheric neutrino analysis.  
Fig.~\ref{tdiff} shows the time-difference distribution between 
the beam spill start and the SK event occurrence after each selection cut.  
%
%
After all cuts we can see a bunch of events in the $1.1 \mu$sec range, 
which corresponds to the beam spill length.  
In the fiducial volume of 22.5 kton, 56 fully contained events 
were observed, which breaks down into 30 single-ring $\mu$-like events, 
2 single-ring e-like events and 24 multi-ring events.  
Fig.~\ref{SKevPOT}(left) shows the SK event number as a function of 
the accumulated POT.  
Fig.~\ref{SKevPOT}(right) shows the distribution of the interval 
between two consecutive 
events.  It fits to an exponential shape quite well, as expected.  
This means that the fluctuation of the event rate is statistical.  
%
%
%
\subsection{Results from event numbers}
A summary of the results is listed in table \ref{sumtab}.  
We observed 56 SK events, while the expectation from the 1kt 
measurement extrapolated by the pion monitor measurement is 
$80.6^{+7.3}_{-8.0}$.  
This means that the probability of the null-oscillation scenario is 
less than 3\%.  \\
We observed 2 single-ring e-like events at SK.  This is consistent 
with the Monte-Carlo expectation, suggesting that these events are 
from the neutral-current $\nu_{\mu}$ interaction with water.  
\begin{table}
\centering
\caption{ \it Summary of the observed and expected numbers of events.
}
\vskip 0.1 in
\begin{tabular}{|l|c|c|c|} \hline
          & observed & expected w/o osci & $\Delta m^{2}=3 \times 10^{-3}$ \\
\hline
\hline
 1-ring $\mu$-like & 30 & $44.0 \pm 6.8$        & 24.4 \\
 1-ring e-like     &  2 & $ 4.4 \pm 1.7$        &  3.7 \\
 multi-ring        & 24 & $32.2 \pm 5.3$        & 24.3 \\
\hline
 total             & 56 & $80.6 ^{+7.3}_{-8.0}$ & 52.4 \\
\hline
\end{tabular}
\label{sumtab}
\end{table}
\subsection{Spectrum analysis}
\begin{figure}[bhtp]
\epsfig{file=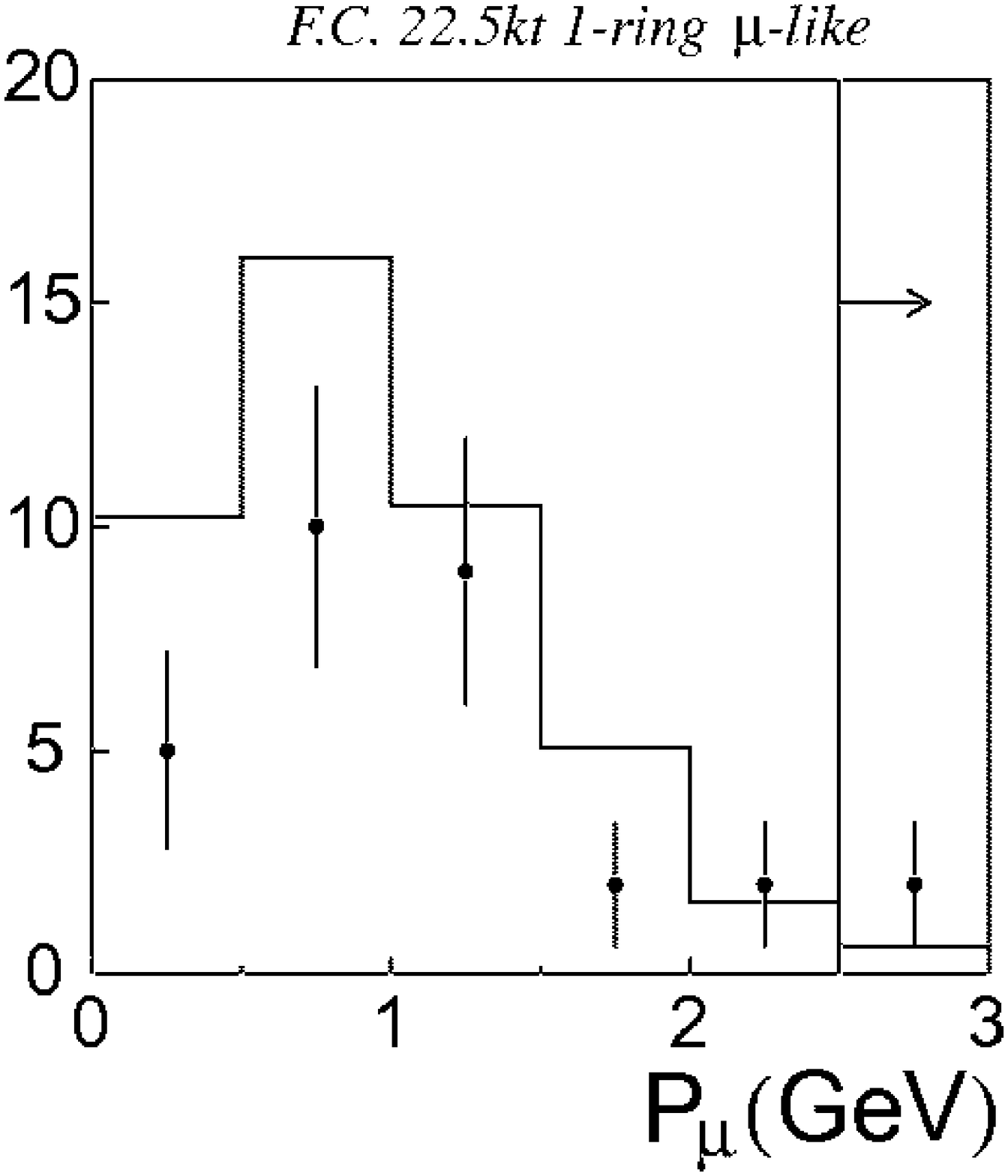,width=7cm}
\hspace{0.5cm}
\epsfig{file=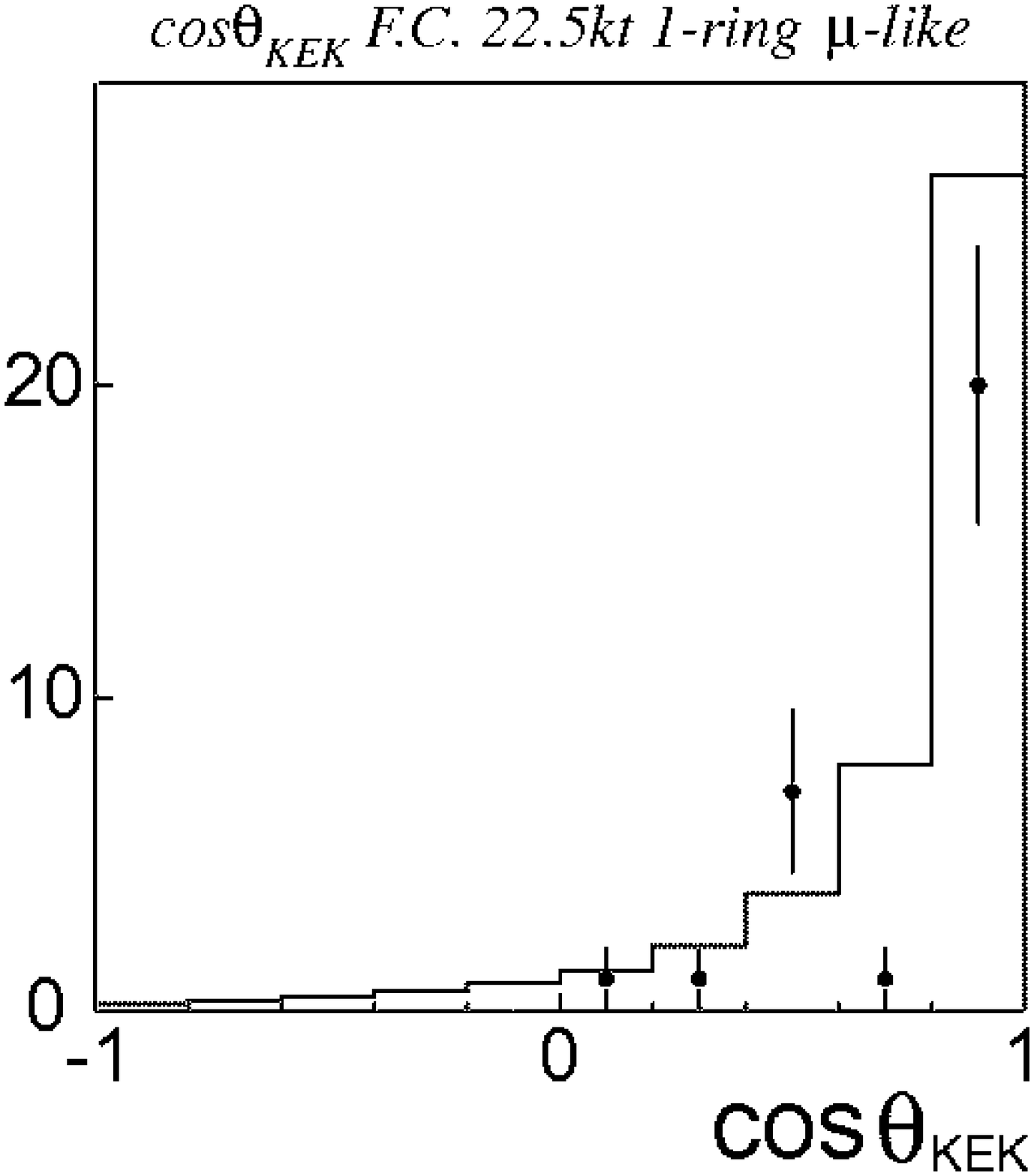,width=7cm}
 \caption{\it
      (left) Muon momentum
      of single-ring $\mu$-like events; 
      (right) Muon angle with respect to the KEK direction 
      of single-ring $\mu$-like events.  
    \label{muangle} }
\end{figure}

\begin{figure}[bhtp]
\centerline{\epsfig{file=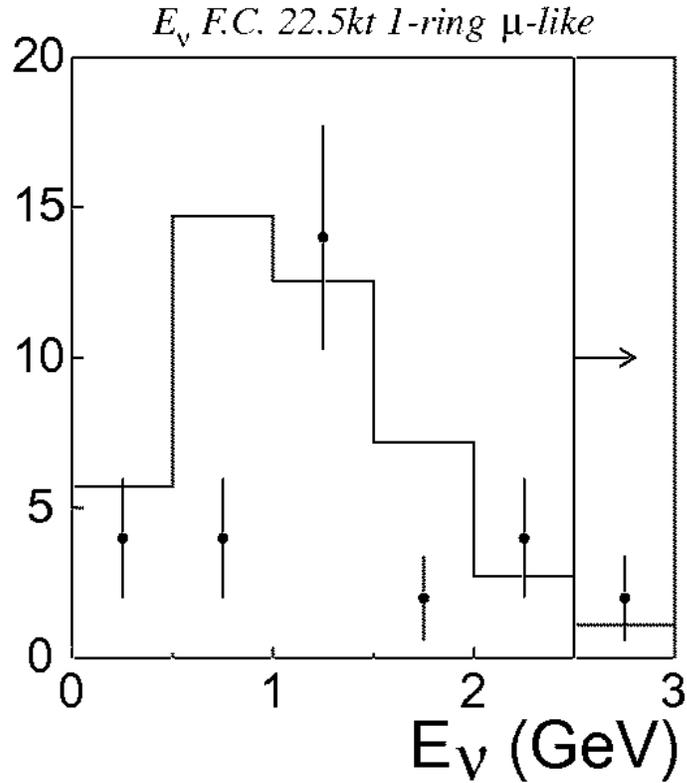,width=9cm}}
 \caption{\it
      Reconstructed neutrino energy of single-ring $\mu$-like events.
    \label{recenu} }
\end{figure}

For a quasi-elastic event, $\nu_{\mu} + n \rightarrow \mu^{-} + p$, 
one can calculate the neutrino energy from the muon energy and angle by 
the following equation:  
\begin{equation}
E_{\nu} = \frac{m_{N}E_{\mu} - m_{\mu}^{2}/2}
           {m_{N}-E_{\mu}+p_{\mu}\cos\theta_{\mu}} . 
\label{eqrecenu} 
\end{equation}
For SK events, in order to enhance the quasi-elastic events, 
we select single-ring $\mu$-like events.  
Fig.~\ref{muangle} shows the muon momentum  and angular distributions.  
Fig.~\ref{recenu} shows the reconstructed 
$E_{\nu}$ distribution calculated by equation~\ref{eqrecenu}.  
The histograms in the figures show the tentative Monte-Carlo calculation 
assuming no oscillation.  A derivation of the spectrum expectation 
from the near-site measurements is under way while taking into account of 
the correct error correlation and the ambiguity of the neutrino interaction.  
%
%
%
%
\section{Near-detector upgrade}
\begin{figure}[bhtp]
\centerline{\epsfig{file=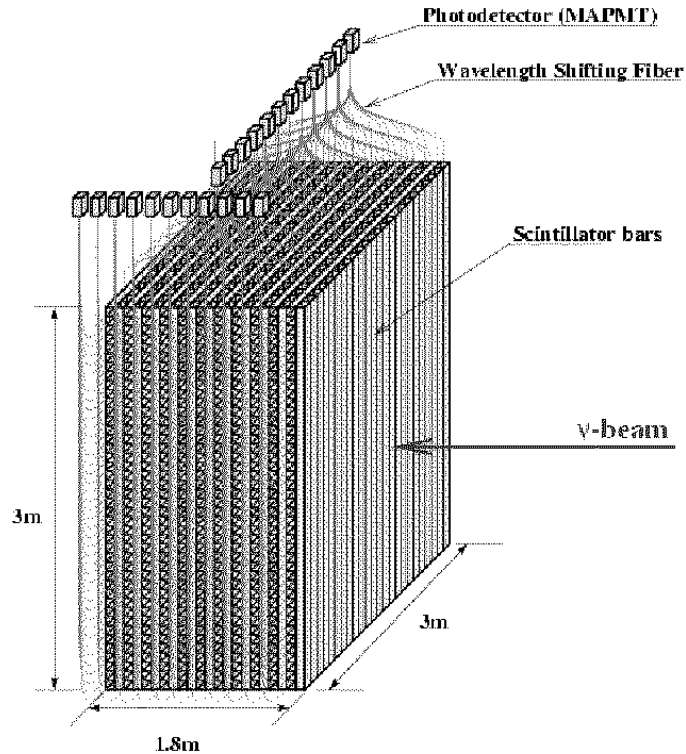,width=9cm}}
 \caption{\it
      Scintillator-bar detector (SCIBAR) to be installed.  
    \label{scibar} }
\end{figure}

If $\Delta m^{2} \approx 3 \times 10^{-3} \rm eV^{2}$, 
as is indicated by atmospheric 
neutrino analyses, the first oscillation minimum 
appears at around $E_{\nu} \approx 0.6$GeV in the K2K experiment.  
It is desirable to monitor 
the neutrino energy spectrum down to this low-energy region at the near site. 
For this purpose, we are going to install a full-active 
scintillator-bar tracker (SCIBAR). 
SCIBAR will replace the lead-glass counters.  
As shown in fig.~\ref{scibar}, 
the detector is made of fine segmented scintillator bars. 
It can detect a proton down to 350 MeV/c.  
Protons can be identified by a dE/dx measurement.  
%
%
%
\section{Run plan}
In order to firmly establish neutrino oscillation, K2K will 
accumulate $10^{20}$ POT, which corresponds to additional 18-month runs.  
In spite of the unfortunate accident of November 2001, 
the SK detector is being rebuilt while aiming at a restart 
within one year.\cite{accident}  
The upgraded detector SCIBAR will be ready by the summer of 2003.  
\section{Conclusion}
During the first half of the K2K run until summer 2001, 
data corresponding to 
$4.8 \times 10^{19}$ POT were accumulated for analyses.  
In this data set, 56 accelerator-produced neutrino 
events were observed in the far detector, 
while the expected number from the near site measurements is 
$80.6^{+7.3}_{-8.0}$.  
This means that the probability of the null-oscillation scenario is 
limited to less than 3\%.  \\
We can declare that a method for long-baseline neutrino experiments 
has been established.  Namely, beam monitoring and handling towards 
the detector 250 km away, synchronization of the far detector 
with the accelerator by the GPS and spectrum and flux extrapolation 
from the near site to the far site.  \\
An oscillation analysis by studying near and far spectra is under way.  
\section{Acknowledgements}
We thank the KEK and ICRR Directorates for their strong support and 
encouragement.  K2K is made possible by the inventiveness  and 
the diligent efforts of the KEK-PS machine group.  We gratefully 
acknowledge the cooperation of the Kamioka Mining and Smelting Company.  
This work has been supported by the Ministry of Education, Culture, 
Sports, Science and Technology, Government of Japan and its grants 
for Scientific Research, the Japan Society of Promotion of Science, 
the U.S. Department of energy, the Korea Research Foundation, and 
the Korea Science and Engineering Foundation.  
\end{document}